\documentclass[11pt]{article}
\usepackage[top = 2 cm, bottom = 2 cm, left = 2.5 cm, right = 2 cm]{geometry}
\usepackage{authblk, cite, color, amssymb, amsmath, enumitem}
\usepackage[hypertex, colorlinks = true, linkcolor = blue, citecolor = red]{hyperref}
\usepackage[dvipsnames]{xcolor}

\begin{document}

\title{Dark Energy from Entanglements with Mirror Universe}

\author[1,2]{M. Gogberashvili}
\author[1]{T. Tsiskaridze}
\affil[1]{\small Javakhishvili State University, 3 Chavchavadze Ave., Tbilisi 0179, Georgia}
\affil[2]{\small Andronikashvili Institute of Physics, 6 Tamarashvili St., Tbilisi 0177, Georgia}

\maketitle

\begin{abstract}

We investigate a possible resolution of the dark energy problem within a pair-universe framework, in which the Universe emerges as an entangled pair of time-reversed sectors. In this setting, a global zero-energy condition allows vacuum energy contributions from the two sectors to cancel, alleviating the need for extreme fine-tuning. We propose that the observed dark energy does not originate from vacuum fluctuations but instead arises as an effective entanglement energy between the visible universe and its mirror counterpart. Treating the cosmological constant as an integration constant fixed by boundary conditions rather than a fundamental parameter, we show that the cosmological equations can be formulated without explicitly introducing vacuum energy. By imposing physically motivated boundary conditions at the cosmological event horizon, we obtain an integration constant consistent with the observed dark energy density. The parallel mirror world scenario thus provides a unified framework that may simultaneously explain the origins of dark energy and dark matter.



\end{abstract}


The cosmological constant problem remains one of the most profound and persistent puzzles in modern theoretical physics \cite{Burgess:2013ara, Ng:1991ri, Weinberg:1988cp}. The cosmological constant, commonly identified with dark energy, is usually interpreted as the energy density of the vacuum. In quantum field theory (QFT), however, vacuum zero-point fluctuations generate a nonvanishing energy density that gravitates and therefore contributes directly to the cosmological constant. Straightforward estimates of this contribution exceed the observed value by many orders of magnitude. Although one may formally introduce a bare cosmological constant of opposite sign to cancel the vacuum contribution, such a cancellation requires extreme fine-tuning between physically independent quantities, raising severe naturalness concerns. This discrepancy strongly suggests that the principles of QFT cannot be naively extrapolated to gravitational and cosmological scales, and that a more fundamental framework—treating gravity, quantum mechanics, and global constraints on an equal footing—is required. Possible directions include quantum gravity, emergent spacetime, or new fundamental symmetries.

In this paper, we propose a resolution of the dark energy problem within a pair-universe framework, motivated by the idea that a quantum universe may emerge from nothing as an entangled pair of universes evolving with opposite directions of time \cite{Boyle:2018tzc, Robles-Perez:2019wll, Zalialiutdinov:2022odv}. In such a construction, the observed universe is accompanied by a mirror counterpart, and the combined system forms a globally symmetric and energetically balanced whole.

The motivation for introducing a pair-universe scenario is closely related to the status of discrete symmetries. In our Universe, the discrete symmetries of charge conjugation ($C$), parity ($P$), and time reversal ($T$), as well as their bilinear combinations such as $CP$ and $PT$, are known to be only approximate. By contrast, the combined $CPT$ symmetry is widely believed to be exact, and any violation would signal physics beyond the Standard Model (SM). In even-dimensional Euclidean spacetime, the inversion $x^\nu \rightarrow -x^\nu$ (where Greek letter indexes take values 0 for temporal and 1, 2, and 3 for spatial components) corresponds to a proper rotation, which upon analytic continuation to pseudo-Euclidean (Lorentzian) spacetime is naturally identified with a $CPT$ transformation \cite{Be-Li-Pi, Lehnert:2016zym}. While $C$, $P$, and $T$ individually belong to the extended Lorentz group, they lack the simple geometric interpretation associated with continuous spacetime rotations.

Among the discrete  symmetries, time reversal $T$ presents the deepest conceptual challenges. In QFT, $T$ operator must be defined as antiunitary in order to preserve the sign of the energy term in the phase factor $e^{iEt}$ of particle wavefunctions when time is reversed. Moreover, causality forbids direct geometric rotations that would take the time axis outside the light cone. As a result, time reversal is implemented algebraically through the interchange of initial and final states in scattering amplitudes \cite{Weinberg}. This construction lacks a clear geometric meaning in curved spacetime and becomes ambiguous near cosmological singularities, where asymptotic states are ill-defined. In addition, particle wavefunctions cannot be eigenstates of the time-reversal operator, in sharp contrast to the case of spatial parity $P$.

These complications motivate an alternative, geometric interpretation of time reversal that avoids negative-energy solutions and does not rely on the interchange of in- and out-states. Since time and space form a unified geometric structure in general relativity, within the pair-universe framework crossing the Big Bang can be viewed as an analytic continuation in which all spacetime coordinates $x^\nu$ change sign. This continuation naturally produces a mirror universe on the opposite side of $t = 0$ time, related to our own by the isometry $x^\nu \to -x^\nu$, corresponding to a $PT$ transformation.

A related point concerns the meaning of the “initial time” ($t=0$) in the pair-universe picture. In the present paper, $t=0$ should not be interpreted as a sharply defined classical instant or as a physical singularity. Rather, it denotes the onset of the semiclassical evolution of each branch after the quantum creation of the entangled universe pair. In quantum cosmology, the emergence of classical spacetime is expected to occur only after decoherence and coarse-graining, so that the notion of a well-defined cosmic time becomes meaningful only beyond this transition.

It is worthy to emphasize that $P$-symmetry is not an invariance of the SM in our universe. Under the conventional parity operation, left-handed  (L) and right-handed (R) spinors transform as $\psi_L \leftrightarrow \psi_R$. However, only left-handed fermions participate in weak charged-current interactions mediated by the $W^\pm$ bosons. Exact parity symmetry can be restored in the two-universe scenario by generalizing the parity transformation to $\psi_L \leftrightarrow \psi'_R$ and $\psi_R \leftrightarrow \psi'_L$, where primed quantities correspond to fields in the mirror sector. These transformations also interchange the SM gauge bosons ($\gamma, W, \dots$) with their mirror counterparts ($\gamma', W', \dots$), rendering the full theory invariant. Accordingly, we adopt the following discrete transformations for fermions \cite{Foot:2014mia}:
\begin{equation} \label{transformations}
\begin{split}
&\psi_L \leftrightarrow \psi'_R~, \quad \psi_R \leftrightarrow \psi'_L~, \quad \text{(parity)} \\
&\psi_L \leftrightarrow \psi'_L~, \quad \psi_R \leftrightarrow \psi'_R~, \quad \text{(time reflection)} \\
&\psi_L \leftrightarrow \psi^*_R~, \quad \psi'_L \leftrightarrow \psi'^*_R~, \quad \text{(charge conjugation)}.
\end{split}
\end{equation}
Here, $T$ and $P$ transformations relate particles across the two sectors, while $C$ acts independently within each sector.

Because parity interchanges SM fermions with mirror fermions of opposite chirality, the mirror universe is naturally identified with the Mirror World \cite{Nishijima:1965zza, Kobzarev:1966qya, Blinnikov:1983gh, Blinnikov:1982eh, Khlopov:1989fj}. Originally introduced to restore left–right symmetry after the discovery of $CP$ violation, mirror-world models have since evolved into a well-developed framework with rich phenomenological implications (see reviews \cite{Berezhiani:2003xm, Okun:2006eb}). In the present construction, the SM is extended by an isomorphic mirror sector, denoted SM$'$, with opposite parity-reflection properties. The extended Lorentz group, including reflections, becomes an exact symmetry of the combined system, ensuring global $CPT$ invariance. The transformations in \eqref{transformations} uniquely preserve the full symmetry of SM and SM$'$ \cite{Foot:2014mia}, while naturally distinguishing left- and right-handed coordinate systems in odd-dimensional space.

Within the paired-universe picture, the pre–Big Bang universe is interpreted as the $PT$-reflected counterpart of our own. Particles whose spacetime coordinates are reversed relative to ours are identified as belonging to the mirror sector. IAs a result, unlike standard approach, time reversal does not require an operator acting within a single Hilbert space. Instead, an additional discrete $Z_2$ symmetry interchanges the two universes and is interpreted as a microscopic spacetime-reversal operation involving both time reversal and the interchange of left- and right-handed coordinate systems.

Because the metric tensor is invariant under coordinate reflections, the two universes may interact gravitationally even in the absence of direct non-gravitational couplings. Such mirror-universe constructions not only restore parity symmetry at a fundamental level but also provide a natural framework for dark matter, identified with matter residing in the hidden sector and interacting predominantly through gravity \cite{Berezhiani:2003xm, Okun:2006eb}. In this scenario, both ordinary and mirror baryons may serve as dark-matter components in their respective universes. Their relative abundances depend on inflaton decay channels, expansion rates, and temperatures of the two sectors, and need not be identical if inflaton and anti-inflaton decay asymmetrically. As a result, baryon densities -- and hence dark matter abundances -- may differ by factors of order unity, making mirror matter a viable hidden-sector dark matter candidate \cite{Berezhiani:2005ek}.

To illustrate the necessity of the pair-universe approach for a consistent definition of $T$ and $CPT$ transformations at the Big Bang, we consider, as a toy model, a real inflaton field $\varphi$ described by the Lagrangian density
\begin{equation} \label{L}
\mathcal{L} = \frac{1}{2} g^{\mu\nu} \partial_\mu \varphi \, \partial_\nu \varphi - \frac{1}{2} \left( m^2 + \frac{R}{6} \right) \varphi^2~,
\end{equation}
where $\partial_\nu \equiv \partial /\partial x_\nu$, $m$ is the inflaton mass and the term $R/6$ represents conformal coupling to gravity. Although $CPT$ symmetry is assumed to hold globally for the universe pair, it may be locally violated within each universe near the cosmological singularity. In this regime, a $CP$ transformation alone is insufficient to map particles to antiparticles, and even neutral fields may become distinguishable from their antiparticle counterparts. This phenomenon is relevant only during the earliest stages of cosmic evolution, when the inflaton field $\varphi(x^\nu)$ governs particle production during reheating in each universe.

During this early epoch, the spacetime geometry may be approximated by a conformally flat metric,
\begin{equation} \label{metric}
g_{\mu\nu} = a^2(\tau) \eta_{\mu\nu}~,
\end{equation}
where $a(\tau)$ is the cosmological scale factor, $\eta_{\mu\nu}$ is the Minkowski metric, and $\tau = \int dt/a(t)$ denotes conformal time. The equation of motion derived from Eq. \eqref{L} then takes the form
\begin{equation}
\frac{1}{a^4} \left( a^2 \dot{\varphi} \right)^{\cdot} + \left( m^2 + \frac{\ddot{a}}{a^3} \right) \varphi = 0~,
\end{equation}
where overdots denote derivatives with respect to conformal time $\tau$. Introducing the field redefinition
\begin{equation} \label{varphi}
\varphi(\tau) = \frac{u(\tau)}{a(\tau)}~,
\end{equation}
reduces the equation to that of a harmonic oscillator with time-dependent frequency,
\begin{equation} \label{u}
\ddot{u} + m^2 a^2(\tau) \, u = 0~.
\end{equation}

For a broad class of scale factors $a(\tau)$, Eq. \eqref{u} reduces to a Bessel equation, admitting oscillatory solutions for large arguments $z = c\tau \gg 1$, where $c$ is an integration constant:
\begin{equation}
J_\alpha(z) \sim \frac{1}{\sqrt{z}} \cos \left( z - \frac{(2\alpha + 1)\pi}{4} \right)~,
\end{equation}
with $\alpha$ denoting the Bessel function order.

Using complex notation for harmonic modes, the inflaton wavefunction may be written as
\begin{equation} \label{u(tau)}
u(\tau) \sim \frac{1}{\sqrt{\tau}} e^{\pm i c \tau}~,
\end{equation}
which permits the standard particle–antiparticle decomposition in terms of creation and annihilation operators. The prefactor $\tau^{-1/2}$ encodes the nontrivial behavior of the inflaton field under time reversal $\tau \to -\tau$.

To highlight the issues associated with time reflection, we now examine two illustrative cosmological backgrounds.

Consider Eq. \eqref{u} for the flat radiation dominated Universe, with the equation-of-state parameter $w = 1/3$, with the scale factor
\begin{equation} \label{scale-radiation}
a (t) \sim t^{\frac {2}{3(1 + w)}} \sim t^{1/2}~,
\end{equation}
which implies that in conformal time $\tau \sim t^{1/2}$,
\begin{equation} \label{scale-tau-radiation}
a (\tau) \sim \tau~.
\end{equation}
Then, the solution to Eq. \eqref{u} can be expressed as
\begin{equation} \label{solution-u-radiation}
u(t) \sim  J_{1/4} \left( ct \right) ~,
\end{equation}
where $c$ is an integration constant. For large arguments, the solution \eqref{solution-u-radiation} exhibits oscillatory behavior, since
\begin{equation}
J_{1/4} \left( ct\right) \simeq \frac {1}{\sqrt t} \, \cos \left(ct - \frac {3\pi}{8} \right)~.
\end{equation}
Thus, the scalar wavefunction \eqref{varphi} can be written as
\begin{equation} \label{varphi-t-radiation}
\varphi (t) \sim \frac {1}{a(t)}\, u(t) \sim \frac {1}{t}\, e^{ic t}~.
\end{equation}
If we assume that the radiation-dominated description, $a(\tau) \sim \tau$, holds right back to the Planck time, the Big Bang singularity arises from the momentary vanishing of the overall conformal factor, $a(\tau) \to 0$, in front of the flat Minkowski metric \eqref{metric}.

A second instructive case is a curvature-dominated universe with equation-of-state parameter $w = -1/3$, corresponding to the boundary between decelerating expansion and inflation. In this case,
\begin{equation} \label{scale}
a (t) \sim t^{\frac {2}{3(1 + w)}} \sim t~,
\end{equation}
and conformal time satisfies
\begin{equation}
\tau \sim \int \frac {dt}{a(t)} \sim \ln t~. \qquad \qquad (t \geq 0)
\end{equation}
Thus,
\begin{equation} \label{scale-tau}
a (\tau) \sim \pm e^{c\tau}~,
\end{equation}
where the negative sign corresponds to time reversal. The assumption of $t\to -t$ symmetry therefore implies a doubling of expansion branches with opposite orientations of $t$ and $\tau$. Equation \eqref{u} then yields
\begin{equation} \label{solution-u}
u(\tau) \sim  J_0 \left( \frac mc \, e^{c\tau}\right) ~,
\end{equation}
where $J_0$ represents the zero-order Bessel functions of the first kind, which for large enough arguments exhibits oscillatory behavior:
\begin{equation}
J_0 \left( \frac mc \, e^{c\tau}\right) \simeq  \sqrt {\frac {2c}{\pi m} \, e^{-c\tau}} \, \cos \left( \frac mc \, e^{c\tau} - \frac \pi 4 \right)~.
\end{equation}
Utilizing the expression for conformal time \eqref{scale-tau}, and using complex notation, one finds that scalar wavefunction \eqref{varphi} behaves as
\begin{equation} \label{varphi-t}
\varphi (t ) \sim \frac {1}{a(\tau)}\, u(\tau) \sim \frac {1}{t^{3/2}}\, e^{imt/c}~.
\end{equation}
Because of the non-analytic factor $t^{3/2}$, time reversal $t \to -t$ cannot be consistently defined for a single real inflaton field. Preservation of time-reversal symmetry at the Big Bang therefore requires a doubling of inflaton states, introducing an anti-inflaton field in the mirror universe,
\begin{equation} \label{varphi'}
\varphi' (t) \sim \frac {1}{t^{3/2}}\, e^{-imt/c}~.
\end{equation}
This mechanism naturally separates the mirror universe from our own prior to reheating.


Another natural question concerns the physical origin of the entanglement between the visible and mirror sectors, and whether such entanglement requires sizable interactions that would contradict cosmological constraints. In the present framework, the entanglement is not assumed to arise from late-time thermal contact between the two universes, but rather from their common quantum origin. In quantum cosmology, when a universe pair is created from “nothing” or through a tunneling/bounce process, the global state is expected to be pure and generically entangled, much like particle–antiparticle pair creation in quantum field theory. In this sense, entanglement is a kinematic consequence of the bipartite origin of the state, rather than the result of sustained microscopic interactions after the sectors have emerged \cite{Robles-Perez:2019wll, Boyle:2018tzc}.

Once the two sectors begin their semiclassical evolution, they quite rapidly decohere from one another, and any non-gravitational cross-couplings must be highly suppressed. This is consistent with the standard mirror-matter cosmology, where the hidden sector is allowed to have a lower temperature than the visible one, in order to satisfy bounds from Big-Bang nucleosynthesis, the cosmic microwave background, and limits on extra relativistic degrees of freedom \cite{Berezhiani:2003xm, Okun:2006eb}. Such constraints imply that any interactions capable of thermalizing the two sectors must be out of equilibrium well before nucleosynthesis, leaving gravity as the dominant universal coupling between them.

Importantly, quantum entanglement does not imply thermal equilibrium. A bipartite pure state can exhibit substantial entanglement entropy even when the two subsystems subsequently evolve with different temperatures and without further energy exchange. The entanglement considered here should therefore be understood as a primordial correlation encoded in the initial quantum state of the universe pair, while the later thermodynamic asymmetry between the sectors reflects their effectively decoupled evolution. Thus, the present scenario remains compatible with strong observational constraints on mirror-sector interactions: entanglement is generated at the birth of the paired universes, while subsequent cross-sector couplings are negligible, preventing thermalization and naturally allowing the two sectors to maintain different temperatures throughout cosmological history.


A key feature of the pair-universe picture is that the visible and hidden sectors act as environments for one another, generating entropy through their mutual entanglement. If the universe pair is regarded as a closed quantum system, its total entropy can be taken to vanish \cite{Landauer}. Under this assumption, entropy production in one universe is exactly compensated by entropy production in its mirror counterpart. By the first law of thermodynamics, such entropy neutrality implies a global zero-energy condition, namely that the total density of all forms of energy vanishes when both sectors are considered together \cite{Gogberashvili:2022cam}. Such a condition allows the universe pair to emerge without violating energy conservation and is consistent with long-standing ideas that the Universe may originate from a state of zero total energy \cite{Fey,Haw}.

Within this framework, the cosmological constant problem acquires a natural reinterpretation. A global zero-energy condition permits the cancellation of vacuum energies between the two time-reflected sectors, thereby eliminating their gravitational effect without the need for fine-tuning \cite{Linde:1988ws}. If the universe pair resides in a global pure quantum state, the visible and mirror sectors must necessarily be entangled. We therefore propose that, although vacuum energy does not explicitly enter the cosmological equations governing the expansion of either universe, the observed dark energy can instead be interpreted as an effective manifestation of the entanglement between the ordinary and mirror universes.

Entanglement, as a fundamentally nonlocal quantum correlation, plays a central role in quantum information theory \cite{Entanglement} and has been extensively studied in gravitational and cosmological contexts, including black hole thermodynamics \cite{Mukohyama:1996yi, Bombelli:1986rw} and early-universe cosmology \cite{Lee:2007zq, Gogberashvili:2022ttz}. Entanglement entropy is defined as the von Neumann entropy of the reduced density matrix obtained by tracing over one subsystem of a bipartite quantum state.

In the present scenario, the natural bipartition is between our Universe and its mirror counterpart, which are classically disconnected but quantum-mechanically entangled, with the surface at $t = 0$ serving as the entangling surface. In this case, the entanglement entropy can be defined using the reduced density matrix obtained by tracing out the degrees of freedom of the mirror universe on the cut of a Cauchy surface by the section of the null surface at $t = 0$ \cite{Lee:2007zq, Kumar:2024nhe}.

Entanglement entropy is usually defined in a form consistent with the holographic principle, although it can also be derived directly from quantum field theory without invoking this principle \cite{Muller1995, Srednicki1993}. While an explicit and fully covariant calculation of the energy associated with entanglement entropy in a dynamical spacetime remains highly nontrivial \cite{Casini:2009sr}, it is nevertheless natural to associate an effective entanglement energy, $E_{\rm Ent}$, with the presence of a cosmological horizon. Such a horizon limits the number of degrees of freedom accessible to a given observer and thereby induces entanglement between observable and unobservable regions of spacetime. In a cosmological setting, this entanglement-induced energy is therefore expected to scale with the horizon size $R_{\rm h}$ \cite{Lee:2007zq}.
\begin{equation}
E_{\rm Ent} = \frac{\beta N_{\rm dof}}{\pi \lambda^2} \, R_{\rm h}~,
\end{equation}
where $\beta$ is a dimensionless coefficient that depends on the spin and statistics of the fields, $N_{\rm dof}$ denotes the effective number of degrees of freedom contributing to the entanglement, and $\lambda$ represents a ultra-violet cutoff characterizing the shortest resolvable length scale.

We suggest that this entanglement energy is the origin of the dark energy density within the horizon,
\begin{equation}
\rho_{\rm DE} =  \frac{3\beta N_{\rm dof}}{(2\pi \lambda R_{\rm h})^2}~.
\end{equation}
The pressure of dark energy,
\begin{equation} \label{p}
p_{\rm DE} = - \frac{1}{3 a^2} \,\frac {d}{da} (a^3 \rho_{\rm DE} ) = -\frac 13 \, \frac{d \rho_{\rm DE}}{d \ln a} - \rho_{\rm DE}~,
\end{equation}
can be derived from the Friedmann equations for a perfect fluid, or equivalently from the conservation of the energy–momentum tensor \cite{Huang:2004ai}. This relation indicates that a perfect fluid whose energy increases as the universe expands exhibits negative pressure. One can therefore associate an equation-of-state parameter with the entanglement energy contribution \cite{Lee:2007zq},
\begin{equation}
\omega_{\rm DE} = \frac {p_{\rm DE}}{\rho_{\rm DE}} = -\frac{1}{3}\left(1 + \frac{4\pi \lambda M_{\rm Pl}\sqrt{\Omega_{\rm DE}}}{\sqrt{\beta N_{\rm dof}}}\right)~,
\end{equation}
where $M_{\rm Pl}$ is the Planck mass and $\Omega_{\rm DE}$ denotes the fractional dark-energy density. The equation of state characterising cosmic acceleration arises only within a restricted region of parameter space. Taking $\lambda \simeq M_{\rm Pl}^{-1}$, $\beta \simeq 0.3$ for scalar fields and $\beta \simeq 0.2$ for fermions \cite{Casini:2009sr}, together with $N_{\rm dof} = 2\times 118$, corresponding to the particle content of the SM and and its mirror counterpart SM$'$, one finds that the effective equation of state is $\omega_{\rm DE} = -1/3$ whenever other forms of energy dominate. It can, however, approach the observed value $\omega_{\rm DE} \simeq -1$ in the dark-energy–dominated regime when the horizon scale is identified with the cosmological event horizon $R_{\rm h} \to R_e$ \cite{Lee:2007zq}.  For these parameter values, we estimate the dark energy density parameter as
\begin{equation} \label{Omega}
\Omega_{\rm DE} = \frac {\rho_{\rm DE}}{\rho_c} = \frac {\beta N_{\rm dof}R_H^2}{(2\pi R_e)^2} \sim \frac {R_H^2}{R_e^2}~,
\end{equation}
where $\rho_c$ denotes the critical density and $R_H \equiv H_0^{-1}$ is the Hubble radius ($H_0$ is the current Hubble parameter). For the present Universe, this expression yields a value of $\Omega_{\rm DE}$ very close to the observed dark energy fraction.

These results support the interpretation of dark energy as an emergent, horizon-related phenomenon arising from quantum correlations, rather than as a fundamental vacuum energy density. At the same time, the sensitivity to the choice of cutoff, horizon definition, and field content highlights the intrinsic difficulties in obtaining precise quantitative predictions within this framework.

An alternative, yet complementary, perspective arises in frameworks where the cosmological constant is not treated as a fundamental parameter of the action, but instead appears as an integration constant determined by global or boundary conditions. This viewpoint is realized, for example, in thermodynamic approaches to gravity \cite{Komatsu:2018meb, Jacobson:1995ab, Padmanabhan:2009vy, Gogberashvili:2016llo}. In such formulations, vacuum energy does not gravitate directly; rather, the accelerated expansion of the Universe is attributed to global constraints on spacetime dynamics, entropy balance, or horizon thermodynamics.

Indeed, combining the Friedmann equation with local energy–momentum conservation for matter leads to cosmological evolution equations that do not explicitly involve a cosmological constant \cite{Gogberashvili:2022cam},
\begin{equation}
\begin{split}
\dot H &= - 4\pi G(\rho + p)~, \\
\dot\rho &= - 3H(\rho + p)~,
\end{split}
\end{equation}
where $\rho$ denotes the total energy density of baryonic and dark matter and $p$ their pressure. Eliminating the combination $(\rho + p)$ yields
\begin{equation}
\frac{4\pi G}{3}\dot\rho = H\dot H~,
\end{equation}
which integrates straightforwardly to
\begin{equation} \label{Hubble}
H^2 = \frac{8\pi G}{3}\rho + C~,
\end{equation}
with $C$ appearing as an integration constant. In this formulation, the cosmological constant is not a parameter of the fundamental action and is therefore free from radiative instability; instead, its value is fixed by boundary conditions imposed on the cosmological solution.

Quantum fluctuations of matter fields should be constrained by the presence of a physical spacetime boundary, which can be naturally identified with the cosmological event horizon $R_e$ \cite{Gaztanaga:2021bgb, Gaztanaga:2022ktb}. In our view, the event horizon -- representing the ultimate causal limit -- is more relevant than the Hubble (apparent) horizon for describing the boundary of quantum particles that are classically disconnected from their mirror counterparts, with the surface at $t = 0$ serving as the entangling surface. The value of the event horizon at the current cosmic time can be estimated as (see, for example, \cite{Margalef-Bentabol:2012kwa}):
\begin{equation} \label{R_e}
R_e = \frac {1}{H_0}\int_{-1}^0 \frac {dy}{\sqrt{\Omega_m (1 + y)^3 + \Omega_{\rm DE}}} \approx \frac {0.96\, R_H}{\sqrt {\Omega_{\rm DE}}} ~,
\end{equation}
where $\Omega_m$ denotes the matter density fraction. At this boundary, it is physically reasonable to assume the absence of matter degrees of freedom responsible for entanglement with mirror particles,
\begin{equation}
\rho |_{R \to R_e} \to 0~.
\end{equation}
Under this condition, Eq. \eqref{Hubble} implies
\begin{equation}
C = H^2|_{R \to R_e} = \frac{1}{R_e^2}~.
\end{equation}
The resulting contribution to the present expansion rate can therefore be expressed as
\begin{equation}
\frac{C}{H_0^2} = \frac{R_H^2}{R_e^2} \approx 1.08 \, \Omega_{\rm DE}~.
\end{equation}
Remarkably, this estimate shows close numerical agreement with the observed dark-energy density, similarly to entropy-based calculations \eqref{Omega}. This result supports the interpretation of dark energy not as a vacuum energy density subject to radiative corrections, but rather as a boundary-induced or entanglement-related effect associated with the finite extent of the observable Universe.


In conclusion, we have explored a possible resolution of the dark energy problem within a pair-universe framework, in which our Universe is accompanied by a time-reversed mirror counterpart. In this construction, the combined system is globally $CPT$ symmetric and satisfies a zero-energy condition, allowing the dominant vacuum energy contributions of the two sectors to cancel without fine-tuning. As a result, the observed dark energy is not identified with vacuum zero-point fluctuations but is instead interpreted as an effective manifestation of quantum entanglement between the visible and mirror universes.

Assuming that the global state of the universe pair is pure, tracing over the hidden sector naturally generates entanglement entropy and an associated entanglement energy. For reasonable choices of the ultraviolet cutoff and the number of degrees of freedom, this energy behaves as a dark-energy–like component with negative pressure. This establishes a physically motivated connection between cosmic acceleration and quantum information, replacing the conventional attribution of dark energy to vacuum fluctuations with an emergent, correlation-driven origin tied to horizon-scale physics.

We further argued that when the cosmological constant is treated as an integration constant rather than a fundamental parameter of the action, its value can be fixed by global boundary conditions imposed at the cosmological event horizon. Requiring the vanishing of the matter density at this boundary yields an integration constant that quite closely matches the observed dark energy density. In this formulation, the cosmological constant problem is reframed as a problem of boundary-condition selection, while dark energy emerges as an entanglement-induced effect intrinsic to the paired-universe structure.

From an observational standpoint, the present framework is distinguishable from Standard Cosmological Model and other dark energy scenarios precisely because dark energy is not a fundamental vacuum component but an emergent, horizon-related entanglement effect. Consequently, the effective equation-of-state parameter $\omega_{\rm DE}$ need not be exactly equal to $-1$ and may exhibit mild redshift dependence governed by horizon-related quantities and the effective number of degrees of freedom. This behavior contrasts with $\Lambda$CDM, where $\omega_{\rm DE}$ is strictly constant, and with quintessence models, where dark energy dynamics arise from local scalar-field evolution rather than global boundary conditions.

A further distinguishing feature is the intimate connection between dark energy and the choice of cosmological horizon. Since the entanglement energy is tied to horizon scales, different horizon definitions—such as the Hubble, event, or apparent horizon—lead to quantitatively distinct late-time expansion histories. This opens the possibility of testing the framework through precision measurements of the expansion rate $H(z)$ (with $z$ denoting the redshift), baryon acoustic oscillations, and Type Ia supernova observations. Unlike various phenomenological dark-energy models, the present approach constrains the allowed parameter space through quantum-field-theoretic considerations of $\lambda$, $\beta$, and $N_{\rm dof}$, thereby reducing arbitrariness.

The entanglement origin of dark energy may also leave imprints on cosmological perturbations. Since the effective dark energy component arises from nonlocal quantum correlations rather than a local fluid or scalar field, its clustering properties are expected to differ from those of standard dynamical dark energy. In particular, the effective sound speed is expected to be close to unity, suppressing dark energy perturbations on subhorizon scales, while subtle deviations may appear near the horizon scale. Such effects could be probed through large-scale structure surveys, the integrated Sachs–Wolfe effect, and cross-correlations between cosmic microwave background anisotropies and matter distributions.

Finally, the pair-universe structure allows for the possibility of small departures from perfect symmetry between the visible and mirror sectors, for instance due to symmetry breaking during cosmological phase transitions. While such effects are expected to be subleading, they could induce a mild time dependence of the effective dark energy density or correlated late-time anomalies. The observation of such signatures would provide a clear discriminator between the present framework and models in which dark energy is a fundamental, isolated component.

In summary, the pair-universe framework offers a conceptually economical and empirically testable resolution of the cosmological constant problem. By disentangling dark energy from vacuum zero-point contributions and linking cosmic acceleration to global quantum correlations and boundary conditions, it provides a unified perspective in which the cosmological constant problem, dark energy, entropy production, and the arrow of time emerge as interconnected consequences of the entangled and globally constrained quantum state of the universe.



\begin{thebibliography}{99}

\bibitem{Burgess:2013ara} C.~P.~Burgess,
``The Cosmological constant problem: Why it's hard to get dark energy from micro-physics,''
doi: 10.1093/acprof:oso/9780198728856.003.0004
[arXiv: 1309.4133 [hep-th]].

\bibitem{Ng:1991ri} Y.~J.~Ng,
``The Cosmological constant problem,''
Int. J. Mod. Phys. D \textbf{1} (1992) 145,
doi: 10.1142/S0218271892000069

\bibitem{Weinberg:1988cp} S.~Weinberg,
``The Cosmological constant problem,''
Rev. Mod. Phys. \textbf{61} (1989) 1,
doi: 10.1103/RevModPhys.61.1

\bibitem{Boyle:2018tzc} L.~Boyle, K.~Finn and N.~Turok,
``CPT-symmetric Universe,''
Phys. Rev. Lett. \textbf{121} (2018) 251301,
doi: 10.1103/PhysRevLett.121.251301
[arXiv: 1803.08928 [hep-ph]].

\bibitem{Robles-Perez:2019wll} S.~J.~Robles-P\'erez,
``Time reversal symmetry in cosmology and the quantum creation of universes,''
Universe \textbf{5} (2019) 150,
doi: 10.3390/universe5060150
[arXiv: 1901.03387 [gr-qc]].

\bibitem{Zalialiutdinov:2022odv} T.~Zalialiutdinov, D.~Solovyev, D.~Chubukov, S.~Chekhovskoi and L.~Labzowsky,
``Alternative interpretation of relativistic time-reversal and the time arrow,''
Phys. Rev. Res. \textbf{4} (2022) L022052,
doi: 10.1103/PhysRevResearch.4.L022052
[arXiv: 2205.13417 [physics.gen-ph]].

\bibitem{Be-Li-Pi} V.~B.~Berestetskii, E.~M.~Lifshitz and L.~P.~Pitaevskii,
                  {\it Quantum Electrodynamics: Vol. 4}
                  (Butterworth-Heinemann, Oxford U.K. 1996).

\bibitem{Lehnert:2016zym} R.~Lehnert,
``CPT symmetry and its violation,''
Symmetry \textbf{8} (2016) 114,
doi: 10.3390/sym8110114.

\bibitem{Weinberg} S.~Weinberg,
                  {\it The Quantum Theory of Fields: Vol. 1}
                  (Cambridge University Press, Cambridge U.K. 1995).

\bibitem{Foot:2014mia} R.~Foot,
``Mirror dark matter: Cosmology, galaxy structure and direct detection,''
Int. J. Mod. Phys. A \textbf{29} (2014) 1430013,
doi: 10.1142/S0217751X14300130
[arXiv: 1401.3965 [astro-ph.CO]].

\bibitem{Nishijima:1965zza} K.~Nishijima and M.~H.~Saffouri,
``CP Invariance and the shadow universe,''
Phys. Rev. Lett. \textbf{14} (1965) 205,
doi: 10.1103/PhysRevLett.14.205.

\bibitem{Kobzarev:1966qya} I.~Y.~Kobzarev, L.~B.~Okun and I.~Y.~Pomeranchuk,
``On the possibility of experimental observation of mirror particles,''
Sov. J. Nucl. Phys. \textbf{3} (1966 837.

\cite{Blinnikov:1983gh}
\bibitem{Blinnikov:1983gh} S.~I.~Blinnikov and M.~Khlopov,
``Possible astronomical effects of mirror particles,''
Sov. Astron. \textbf{27} (1983) 371.

\bibitem{Blinnikov:1982eh} S.~I.~Blinnikov and M.~Y.~Khlopov,
``On possible effects of 'miror' particles,''
Sov. J. Nucl. Phys. \textbf{36} (1982) 472.

\bibitem{Khlopov:1989fj} M.~Y.~Khlopov, G.~M.~Beskin, N.~E.~Bochkarev, L.~A.~Pustylnik and S.~A.~Pustylnik,
``Observational physics of Mirror World,''
Sov. Astron. \textbf{35} (1991) 21.

\bibitem{Berezhiani:2003xm} Z.~Berezhiani,
``Mirror world and its cosmological consequences,''
Int. J. Mod. Phys. A \textbf{19} (2004) 3775,
doi: 10.1142/S0217751X04020075
[arXiv: hep-ph/0312335].

\bibitem{Okun:2006eb} L.~B.~Okun,
``Mirror particles and mirror matter: 50 years of speculations and search,''
Phys. Usp. \textbf{50} (2007) 380,
doi: 10.1070/PU2007v050n04ABEH006227
[arXiv: hep-ph/0606202].

\bibitem{Berezhiani:2005ek} Z.~Berezhiani,
``Through the looking-glass: Alice's adventures in mirror world,''
doi: 10.1142/9789812775344\_0055
[arXiv: hep-ph/0508233].

\bibitem{Landauer} R.~Landauer,
``Information is physical,''
Phys. Today \textbf{44} (1991) 23,
doi: 10.1063/1.881299.

\bibitem{Gogberashvili:2022cam} M.~Gogberashvili,
``Towards an information description of space-time,''
Found. Phys. \textbf{52} (2022) 74,
doi: 10.1007/s10701-022-00594-6
[arXiv: 2208.13738 [physics.gen-ph]].

\bibitem{Fey} R.P. Feynman, F.B. Morinigo and G. Wagner,
             {\it Feynman Lectures on Gravitation}
             (Addison-Wesley, Reading 1995).

\bibitem{Haw} S. Hawking,
               {\it A Brief History of Time}
               (Bantam, Toronto 1988).

\bibitem{Linde:1988ws} A.~D.~Linde,
``The Universe multiplication and the cosmological constant problem,''
Phys. Lett. B \textbf{200} (1988) 272,
doi: 10.1016/0370-2693(88)90770-8

\bibitem{Entanglement} M.~A.~Nielsen and I.~L.~Chuang,
{\it Quantum Computation and Quantum Information}
(Cambridge University Press, Cambridge, 2001).

\bibitem{Mukohyama:1996yi} S.~Mukohyama, M.~Seriu and H.~Kodama,
``Can the entanglement entropy be the origin of black hole entropy?,''
Phys. Rev. D \textbf{55} (1997) 7666,
doi: 10.1103/PhysRevD.55.7666
[arXiv: gr-qc/9701059 [gr-qc]].

\bibitem{Bombelli:1986rw} L.~Bombelli, R.~K.~Koul, J.~Lee and R.~D.~Sorkin,
``A Quantum source of entropy for black holes,''
Phys. Rev. D \textbf{34} (1986) 373,
doi: 10.1103/PhysRevD.34.373

\bibitem{Lee:2007zq} J.~W.~Lee, J.~Lee and H.~C.~Kim,
``Dark energy from vacuum entanglement,''
JCAP \textbf{08} (2007) 005,
doi: 10.1088/1475-7516/2007/08/005
[arXiv: hep-th/0701199 [hep-th]].

\bibitem{Kumar:2024nhe} N.~Kumar,
``On the accelerated expansion of the universe,''
Grav. Cosmol. \textbf{30} (2024) 85,
doi: 10.1134/S0202289324010080
[arXiv: 2406.04392 [gr-qc]].

\bibitem{Muller1995} R.~Muller and C.~O.~Lousto,
``Entanglement entropy in curved space-times with event horizons,''
Phys. Rev. D \textbf{52} (1995) 4512,
doi: 10.1103/PhysRevD.52.4512
[arXiv: gr-qc/9504049 [gr-qc]].

\bibitem{Srednicki1993} M.~Srednicki,
``Entropy and area,''
Phys. Rev. Lett. \textbf{71} (1993) 666,
doi: 10.1103/PhysRevLett.71.666
[arXiv: hep-th/9303048 [hep-th]].

\bibitem{Gogberashvili:2022ttz} M.~Gogberashvili,
``Fixing cosmological constant on the event horizon,''
Eur. Phys. J. C \textbf{82} (2022) 1049,
doi: 10.1140/epjc/s10052-022-11033-1
[arXiv: 2301.04334 [gr-qc]].

\bibitem{Casini:2009sr} H.~Casini and M.~Huerta,
``Entanglement entropy in free quantum field theory,''
J. Phys. A \textbf{42} (2009) 504007,
doi: 10.1088/1751-8113/42/50/504007
[arXiv: 0905.2562 [hep-th]].

\bibitem{Huang:2004ai} Q.~G.~Huang and M.~Li,
``The Holographic dark energy in a non-flat universe,''
JCAP \textbf{08} (2004) 013,
doi: 10.1088/1475-7516/2004/08/013
[arXiv: astro-ph/0404229 [astro-ph]].

\bibitem{Komatsu:2018meb} N.~Komatsu,
``Generalized thermodynamic constraints on holographic-principle-based cosmological scenarios,''
Phys. Rev. D \textbf{99} (2019) 043523,
doi: 10.1103/PhysRevD.99.043523
[arXiv: 1810.11138 [gr-qc]].

\bibitem{Jacobson:1995ab} T.~Jacobson,
``Thermodynamics of space-time: The Einstein equation of state,''
Phys. Rev. Lett. \textbf{75} (1995) 1260,
doi: 10.1103/PhysRevLett.75.1260
[arXiv: gr-qc/9504004].

\bibitem{Padmanabhan:2009vy} T.~Padmanabhan,
``Thermodynamical aspects of gravity: New insights,''
Rept. Prog. Phys. \textbf{73} (2010) 046901,
doi: 10.1088/0034-4885/73/4/046901
[arXiv: 0911.5004 [gr-qc]].

\bibitem{Gogberashvili:2016llo} M.~Gogberashvili and U.~Chutkerashvili,
``Cosmological constant in the thermodynamic models of gravity,''
Theor. Phys. \textbf{2} (2017) 163,
doi: 10.22606/tp.2017.24002
[arXiv: 1605.04197 [physics.gen-ph]].

\bibitem{Gaztanaga:2021bgb} E.~Gaztanaga,
``The cosmological constant as a zero action boundary,''
Mon. Not. Roy. Astron. Soc. \textbf{502} (2021) 436,
doi: 10.1093/mnras/stab056
[arXiv: 2101.07368 [gr-qc]].

\bibitem{Gaztanaga:2022ktb} E.~Gaztanaga,
``The cosmological constant as event horizon,''
Symmetry \textbf{14} (2022) 300,
doi: 10.3390/sym14020300
[arXiv: 2202.00641 [astro-ph.CO]].

\bibitem{Margalef-Bentabol:2012kwa} B.~Margalef-Bentabol, J.~Margalef-Bentabol and J.~Cepa,
``Evolution of the cosmological horizons in a concordance universe,''
JCAP \textbf{12} (2012) 035,
doi: 10.1088/1475-7516/2012/12/035
[arXiv: 1302.1609 [astro-ph.CO]].

\end{thebibliography}
\end{document}